\documentclass[useAMS,usenatbib]{mn2e}
\usepackage{psfig}

\def\kms{\relax \ifmmode {\,\rm km\,s}^{-1}\else \,km\,s$^{-1}$\fi}

\def\mincir{\ \raise-2.truept\hbox{\rlap{\hbox{$\sim$}}\raise5.truept
    \hbox{$<$}\ }}
\def\magcir{\ \raise-2.truept\hbox{\rlap{\hbox{$\sim$}}\raise5.truept
    \hbox{$>$}\ }}

\def\arcsec{\hbox{$^{\prime\prime}$}}
\def\nii{[N {\sc ii}]}
\def\sii{[S {\sc ii}]}

\def\oii{[O {\sc ii}]}
\def\oi{[O {\sc i}]}
\def\heii{He{\sc ii}}
\def\hei{He{\sc i}}
\def\oiii{[O {\sc iii}]}

\def\ha{H$\alpha$}
\def\hb{H$\beta$}
\def\ki{$\rm{Knot1}$}
\def\kii{$\rm{Knot2}$}
\def\co{$\rm{Core}$}

\def\chb{$c_{\rm H\beta}$}
\def\te{$T_e$}
\def\teff{$T_{eff}$}
\def\ne{$N_e$}

\title[K~4-47: photons versus shocks]{K~4-47: a planetary nebula
excited by photons and shocks\thanks{Based on observations obtained at
the 2.5~INT telescope of the Isaac Newton Group in the Spanish
Observatorio de Roque de los Muchachos of the the Instituto de
Astrof\'\i sica de Canarias.}}

\author[D. R. Gon\c calves et al.]{D. R. Gon\c calves$^{1}$\thanks{E-mail:
denise@ll.iac.es};
A. Mampaso$^{1}$; R. L. M. Corradi$^{2}$; M. Perinotto$^{3}$;
\newauthor
A. Riera$^{4}$ and L. L\'opez-Mart\'\i n$^{1}$
\\
$^{1}$Instituto de Astrof\'{\i}sica de Canarias, E-38205 La Laguna,
Tenerife,
     Spain\\
$^{2}$Isaac Newton Group of Telescopes, Apartado de Correos 321, E-38700
Sta.
     Cruz de La Palma, Spain\\
$^{3}$Dipartimento di Astronomia e Scienza dello Spazio, Universit\`a di
     Firenze, Largo E. Fermi 5, 50125 Firenze, Italy\\
$^{4}$Departament de F\`\i sica i Enginyeria Nuclear, Universitat
Polit\'ecnica
de Catalunya, Avenida V\'\i ctor Balaguer s/n, \\
E-08800 Vilanova i La Geltr\'u, Spain}

\begin{document}

\date{Accepted ?. Received ?; in original form ?}

\pagerange{\pageref{firstpage}--\pageref{lastpage}} \pubyear{2004}

\maketitle

\label{firstpage}

\begin{abstract}
K~4-47 is an unusual planetary nebula composed of a compact
high-ionization core and a pair of low-ionization knots. Long-slit
medium-resolution spectra of the knots and core are analyzed in this
paper. Assuming photoionization from the central star, we have derived
physical parameters for all the nebular components, and the
({\it{icf}}) chemical abundances of the core, which appear similar to
Type-I PNe for He and N/O but significantly deficient in oxygen.  The
nebula has been further modelled using both photoionization (CLOUDY)
and shock (MAPPINGS) codes.  From the photoionization modelling of the
core, we find that both the strong auroral \oiii~4363\AA\ and
\nii~5755\AA\ emission lines observed and the optical size of the core
cannot be accounted for if a homogeneus density is adopted.  We
suggest that a strong density stratification, matching the
high-density core detected at radio wavelengths and the much lower
density of the optical core, might solve the problem.  From the
bow-shock modelling of the knots, on the other hand, we find that
knots' chemistry is also represented by Type-I PN abundances, and that
they would move with velocities of 250 - 300\kms.
\end{abstract}


\section{Introduction}

Planetary nebulae (PNe) are known to possess a variety of small-scale
structures that are usually in a lower ionization state than the main
body of the nebulae.  The morphological and kinematic properties of
these low-ionization structures (LISs, Gon\c calves, Corradi \&
Mampaso, 2001) vary from type to type, in the sense that LISs can
appear in the form of pairs of knots, filaments, jets, and isolated
features moving with velocities that either do not differ
substantially from that of the ambient nebula, or instead move
supersonically through the environment.

The total number of PNe that are known to possess LISs is 55, i.e.
about 10\% of all the 527 Galactic PNe imaged in filters of high- and
low-ionization emission lines (Balick 1987; Schwarz, Corradi \&
Melnick 1992; Manchado et al.  1996).  The different types of LISs may
be easily seen in Corradi et al.  (1996).

At present, the origin of jets and pairs of knots in PNe is not
completely clear.  From the theoretical point of view, the principal
physical process behind the formation of collimated LISs is the
interplay between the stellar AGB and post-AGB winds (for single
stars) or between stellar and disk winds (if the central star is
binary).  According to the various studies dedicated to this issue (e.g.
Garc\'\i a-Segura~1997; Garc\'\i a-Segura \& L\'opez 2000; Steffen, L\'opez \& Lim 
2001; Blackman, Frank \& Welch 2001) jets and knots originated by
this interplay are predicted to be supersonic, highly collimated and
two-sided. In the case of single stars these are expected to be
produced at the same time as the main PN shell, but in the case of a
binary star origin they may be younger than the host PN.  However,
important properties of these LISs such as density contrasts, the
peculiar nitrogen abundance and main excitation mechanisms, appear
hard to explain (see Dwarkadas \& Balick 1998; Balick \& Frank 2002;
Gon\c calves et al.~2003).

K~4-47 (PN G149.0+04.4) is a compact PN that contains LISs.  It is
composed by a small, high ionization nebular core and a pair of
low-ionization, high-velocity knots, connected to the core by a much
fainter low-ionization lane (Corradi et al.~2000; see Figure 1).
Gon\c calves et al.~(2001) have proposed that the low-ionization lanes
and knots of K~4-47 are genuine jets.  Their morphological and
ki\-ne\-ma\-ti\-cal properties are explainable if the jets and knots were
formed by accretion disks, attaining velocities of several hundred
kilometers per second (the main properties of these models are also
summarized in Gon\c calves et al.~2001 and Balick \& Frank~2002).
These highly supersonic velocities imply that the resulting LISs are
likely to be shock-excited.

K~4-47 is a poorly studied PN, for which statistical methods provide
an unrealistic wide range of distances, for instance, 8.5~kpc (Cahn,
Kaler \& Stanghellini 1992) or 26~kpc (van de Steene \& Zijlstra
1994).  Corradi et al.  (2000) computed a distance between 3~kpc and
7~kpc assuming that the object participates to the ordered rotation of
the disk of the Galaxy, but note that the relatively large height of
K~4-47 on the Galactic plane (0.54~kpc for a distance of 7~kpc) adds
some further uncertainty to this determination.  Tajitsu \& Tamura
(1998) estimated a distance of 5.9~kpc using the integrated IRAS
fluxes under the (crude) assumption of constant dust mass for all PNe.
Lacking of anything better, we shall adopt in the following this
distance of 5.9~kpc.  Lumsden et al.~ (2001) mapped the H$_2$ emission
from K~4-47 finding that it is excited by shocks.  The object also
appears in the 6~cm VLA radio survey of Aaquist \& Kwok (1990) showing
a very compact radio core, with a diameter of 0.25 arcsec, and one of
the largest brightness temperatures (T$_b$=8700 K) found in PNe.  So
far, the properties (luminosity and temperature) of its central star
as well as its nebular (physical and chemical) properties are not
known.

In this work, we address the debated issue of the nature and origin of
high-velocity LISs, through the determination of the physical
parameters, excitation and chemistry of K~4-47.  Spectra of the core
and the pair of knots are analyzed using two different models, which
consider the gas to be either fully photoionized by the PN central
star,  or fully ionized by shocks.  We will show that K~4-47 is
particularly interesting for this study because, in contrast to most
of the LISs studied up to now (Dopita 1997; Dwarkadas \& Balick 1998;
Miranda et al.~2000; Gon\c calves~2003; Gon\c calves et al.~2003), its
pair of LISs is mainly shock excited.

\section{Data Acquisition and Reduction}

Spectra of K~4-47 were obtained on 2001 August 28 at the 2.5m~Isaac
Newton telescope (INT) at the Observatorio del Roque de los Muchachos
on La Palma (Spain), using the Intermediate Dispersion Spectrograph
(IDS).  The IDS was used with the 235~mm camera and the R300V grating,
providing a spectral coverage from 3650 \AA\ to 7000 \AA\ with a
spectral reciprocal dispersion of 3.3 \AA~pix$^{-1}$ and a
resolution of 6.9~\AA.  The spatial
scale of the instrument was 0$''$.70~pix$^{-1}$ with the TEK5 CCD.
Seeing varied from 1$''$.1 to 1$''$.2.  The slit width and length were
1.5\arcsec\ and 4\arcmin, respectively.  The slit was positioned
through the centre of the nebula at P.A.  = 41$^{\circ}$, passing
through the knots, and the exposure times were 3$\times $1800~s.

\begin{figure}
\vbox{
\psfig{file=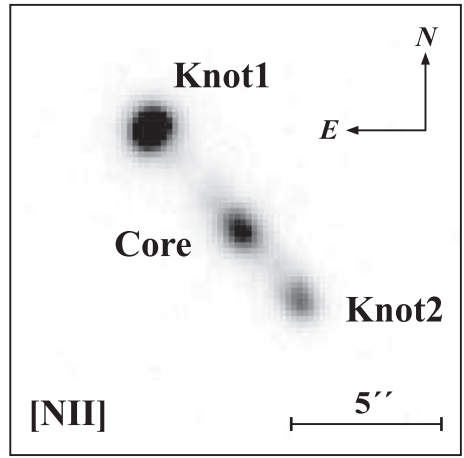,width=6.0truecm,bbllx=199pt,bblly=413pt,bburx=370pt,bbury=584pt}}
\caption{\nii~6583\AA\ image of K~4-47, adapted from Corradi et al. (2000).
Labels of the selected structures are indicated.}
\label{image}
\end{figure}

Bias frames, twilight and tungsten flat-field exposures, arcs and
exposures of standard stars BD+332642, Cyg OB2\#9, HD19445, and
BD+254655, were obtained.  Spectra were reduced and flux calibrated
using the standard IRAF package for long-slit spectra.  Line fluxes
were measured se\-pa\-ra\-te\-ly for the three nebular regions indicated in
Figure~\ref{image}, namely \ki, \co\ and \kii.  The
observed line fluxes are given in Table~1.  Errors in the fluxes were
calculated taking into account the errors in the
measurement of the fluxes, as well as systematic errors of the flux
calibrations, background determination, and sky subtraction.  The
bottom lines of Table~1 give the estimated accuracy of the measured
fluxes for a range of line fluxes (relative to \hb) in each of the
regions.  Absolute \hb\ fluxes integrated along the slit in each
region are as follows: F(\hb)$_{\rm {Knot1}}$=2.41; F(\hb)$_{\rm
{Core}}$=2.76; and F(\hb)$_{\rm {Knot2}}$=1.64 (in units of
$10^{-15}$ erg cm$^{-2}$ s$^{-1}$).
 
\section[]{Physical Parameters and diagnostic diagrams}

\begin{figure}
\vbox{
\psfig{file=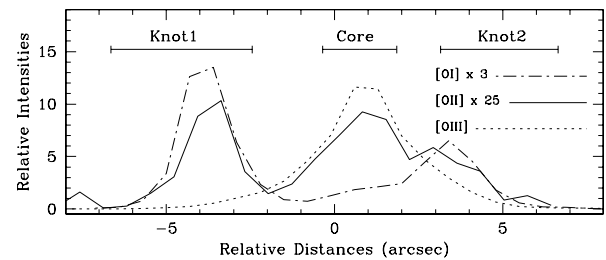,width=8.0truecm,bbllx=90pt,bblly=385pt,bburx=444pt,bbury=544pt}}
\caption{Spatial intensity profile of \oi~6300\AA,
\oii~(3726\AA+3729\AA)\ and \oiii~5007\AA\ emission
lines along P.A.=41$^\circ$.}
\label{ratio}
\end{figure}

Figure~\ref{ratio} shows the spatial profiles of low- to
high-excitation oxygen emission lines (\oi~6300\AA,
\oii~(3726\AA+3729\AA) and \oiii~5007\AA); we clearly see that the
knots have a much lower excitation than the core.  While the
low-ionization profiles (\oi~6300\AA\ and \oii~(3726\AA+3729\AA)) show
local maxima at the positions of the knots, the \oiii~5007\AA\ profile
presents its maxima in the core and does not peak at the knots.  The
average values of \nii~6583\AA/\ha\ at the knots are 2.0 (\ki) and 2.1
(\kii), much higher than those usually found for spherical and
elliptical PNe (e.g., Aller \& Czyzak 1983), although such values are
frequently found in the Peimbert Type I objects (Peimbert 1978;
Peimbert \& Torres-Peimbert 1983).  The \co, on the other hand, has a
less extreme \nii~6583\AA/\ha\ line ratio, 1.2, even though still
higher than in most PNe.  Peculiar line ratios like those found in
K~4-47 have been interpreted as indicative of the presence of
shock-excitation or anomalous abundances (see, for instance, Miranda
\& Solf 1992).

Fluxes were extinction-corrected using \chb\ (the logarithmic ratio
between observed and dereddened \hb\ fluxes), determined from the
observed \ha/\hb\ and \ha/H$\gamma$ ratios, for each of the selected
features of the nebula, and the reddening law of Cardelli, Clayton, \&
Mathis (1989).  The derived \chb\ are: 1.26$\pm$0.23(\ki);
1.37$\pm$0.17(\co); 1.11$\pm$0.17(\kii).  The physical properties (for
all the ne\-bu\-lar regions) and chemical abundances of the \co\ were then
evaluated assuming pure photoionization by the central star.

\subsection{Density and Temperatures}

We estimate the electron densities of K~4-47, which are listed in
Table~1, using the \sii~6717\AA/6731\AA\ line ratio. Both knots are
denser than the core (1900~cm$^{-3}$) by factors of 2.4 and 1.2 for
\ki\ and \kii, respectively.

Adopting these densities, electron temperatures are estimated
using standard line ratios: $T_e$\oiii\ is obtained from the
\oiii~4959\AA/4363\AA\ line ratio, which is appropriate to zones of
medium to high excitation, and $T_e$\nii\ from \nii~6583\AA/5755\AA,
representative of low-excitation regions.  The $T_e$\oiii\ for \ki\
and the $T_e$\nii\ of the \co, are lower limits.  We find that, within
the errors and when both determinations are available, the two
temperatures are similar. Errors on
$T_e$\oiii\ at the position of the knots are significantly higher than
those of the \nii\ temperatures simply because of the low-ionization
nature of these regions.  Note that \te\ is remarkably higher than the
typical values for PNe of around $10^4$~K (Kaler 1986).

\subsection{Diagnostic Diagrams}

\begin{figure}
\vbox{
\psfig{file=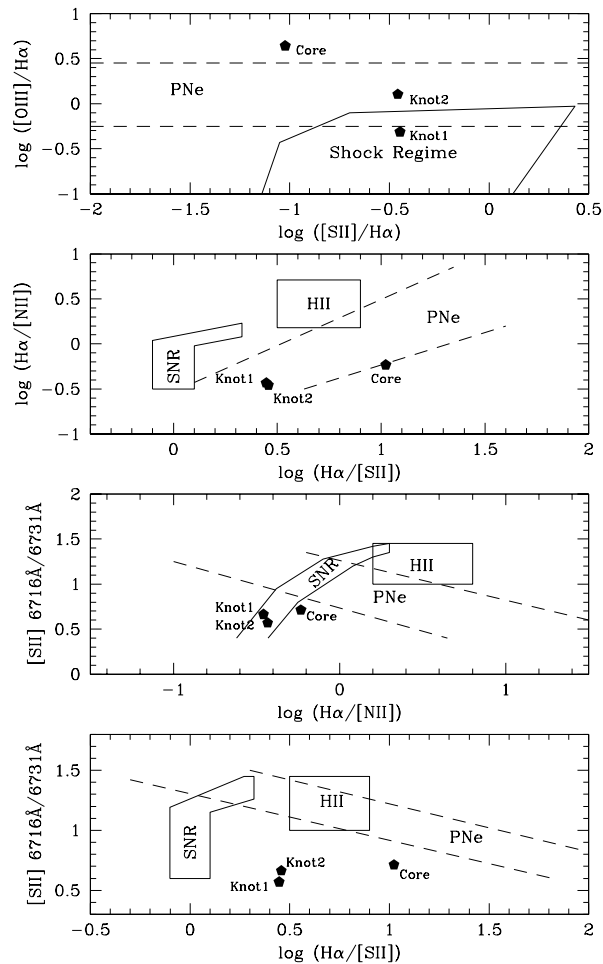,width=8.0truecm,bbllx=120pt,bblly=145pt,bburx=470pt,bbury=715pt}}
\caption{
Diagnostic diagrams showing the loci of the core and knots in K~4-47.
They include the empirical loci
of the PNe, supernova remnants (SNR) and H{\sc ii} regions (HII), as
well as the theo\-re\-ti\-cal ``Shock Regime" zone as defined by Phillips and
Cuesta (1999) using the plane-parallel and bow-shock models of Hartigan,
Raymond \& Hartmann (1987). Line ratios considered are the sum of the
doublets (\oiii$\equiv$\oiii~4959\AA+5007\AA;
\sii$\equiv$\sii~6716\AA+6731\AA;
\nii$\equiv$\nii~6548\AA+6583\AA) when not followed by the specific
wavelength.  
}
\label{shockphot}
\end{figure}

Figure~\ref{shockphot} shows several diagnostic diagrams which are
commonly used to distinguish photoionized nebulae from shock-excited
objects, adapted from Sabbadin, Minello \& Bianchini (1977) and
Phillips \& Cuesta (1999). In all of them, the knots of K~4-47 are
located close to the shock-excited zones while the core is displaced
toward the locus where PNe are usually found, i.e.  photoionized
objects.  This suggests that shock-excitation plays an important role
in the knots, but not in the core. The fact that in the
\oiii~(4959\AA+5007\AA)/\ha\ {\it{vs.}} \sii~(6716\AA+6731\AA)/\ha\
diagram \kii\ do not fall exactly within the ``Shock Regime" area is
not against this interpretation, as its boundaries were originally
computed for young stellar objects, using pa\-ra\-me\-ters (gas abundances
and shock velocities) which can be different from those of evolved
objects like K~4-47.

The large expansion velocities of the knots and their broadened line
profiles (Corradi et al.~2000) are also typical of shocked flows like
H-H objects.  In addition, the shock-excited nature of the knots would
be consistent with their measured $T_e$ (Table 1), unusually large for
low-ionization microstructures in PNe, and with their extremely high
\nii~6583\AA/\ha\ line ratio. On the contrary, the velocities observed
in the core are substantially lower than in the knots, and so is its
\nii~6583\AA/\ha\ line ratio.

It is therefore suggested that the high velocity knots \ki\ and \kii\ of
K~4-47 are mainly shock heated, while the \co\ is mainly photoionized.
With the aid of existing photoionization and shock modelling we will
explore in the following these alternatives.  

\section{Analysis of the core}

\subsection{Empirical Abundances}

Ionic and total abundances for the \co\ were computed using the ionization
correction factors ({\it icf}) scheme, following Kingsburgh \& Barlow
(1994), as described in Corradi et al.  (1997) and Gon\c calves et
al.~(2003).  Results are shown in Table~2. Note that as the $T_e$\nii\
in Table~1 is only a lower limit, the abundances of the low-ionization
ions are upper limits.  Errors on the total abundances are obtained by
pro\-pa\-ga\-ting the errors in the ionic abundances as well as on the {\it
icf}, and are indicated within brackets in the table as percentage
errors (14\% for He, for which no {\it icf} is applied, and 30--50\%
for O and N).

According to its large He and N/O abundances, K~4-47 is similar to
most bipolar PNe (Perinotto \& Corradi 1998), and would be classified
as a Type I PN (Peimbert \& Torres-Peimber~1983, see also Kingsburgh
\& Barlow 1994).  However, the O, Ne and S abundances of K~4-47 are
among the lowest measured for Galactic PNe, and are more typical of
some halo (or Type IV) PNe (see e.g.  Howard, Henry \& McCartney 1997).

High He/H abundances (0.15 $\le$ He/H $\le$ 0.20), and some (but much
less extreme than in K~4-47) oxygen deficiency (1.35 $\le$ O/H
($\times 10^{-4}$) $\le$ 2.7), were found in other bipolar PNe, such
as NGC 6302 (Pottasch \& Beintema 1999), NGC~6537 and He~2-111
(Pottasch, Beintema \& Feibelman 2000).  Very much like in the case of
K~4-47, these PNe have considerably high electron temperatures
(average $T_e$, from optical, ISO and IUE spectra for the three of
them are 17200~K up to 18300~K).  Therefore, we have the somewhat
contradictory result that while K~4-47 seems to be an extreme bipolar
PN in terms of its morphology and He and N abundances, its O, Ne
and S abundances rather resemble PNe in the
Galactic halo, as its relatively large height on the Galactic plane
might also suggest (see Section 1), and contrary to bipolar PNe which
are highly concentrated toward the Galactic plane (Corradi \& Schwarz
1995).

To further investigate these issues, in the next two Sections we will
model the \co\ spectrum using, first, a shock excitation code, and,
then, a pure photoionization model.

\subsection{Plane-parallel Shock Modelling}
 
Shock models have successfully reproduced the emission line ratios,
line profiles and, recently, velocity maps in HH objects (Beck et al.
2004). To check the possible influence of shock excitation on the \co\
spectrum we have explored a variety of steady plane-parallel models,
for which we use the photoionization-shock code MAPPINGS Ic (Dopita, Binette \& 
Tuohy 1984; Binette, Dopita \& Tuohy 1985).

One important characteristic of shock models is that the predicted
spectra strongly depends on the preshock io\-ni\-za\-ti\-on. We have compiled
a set of fully pre-ionized shock models (i.e. the incident gas has
been ionized) and a set of local equlibrium preionization for the gas
entering the shock (as described by Shull \& McKee 1979, Hartigan et
al.  1987).  Relative abundances typical of Type I PNe were assumed,
and two values for the pre-shock density were considered (100, 1000
cm$^{-3}$). The input shock velocities were varied from 95 to 140 km
s$^{-1}$, i.e. the range producing large \oiii~(4959\AA,5007\AA)/\hb\
ratios (Hartigan et al. 1987), as observed. Note, however, that the
measured expansion velocity of the \co\ is smaller than 50\kms\
(Corradi et al. 2000), pointing to lower shock velocities.

None of the predicted spectra could reproduce the large observed
ratios of intermediate to high excitation emission lines with respect
a H {\sc i} Balmer line (i.e. \oii~3727\AA/\hb, \oiii~5007\AA/\hb, [Ne
{\sc iii}]~3868\AA/\hb\ and He {\sc ii}~4686\AA/\hb).  Moreover, the
shocked spectra predicted by our models imply \oi~6300\AA/\ha\ and
\sii~(6716\AA+6731\AA)/\ha\ larger than the observed values. A large
discrepancy was also found with the observed \oiii~4959\AA/4363\AA\
emission line ratio, which is more than 2 times larger than the model
values. And, finally, \nii~6583\AA/5755\AA\ is very high, $\approx$
300, because the intensity of \nii~5755\AA\ is largely understimated
by these models.

We conclude that the plane-parallel shock models are not able to
reproduce the spectrum of the core of K~4-47.  

\subsection{Photoionization Modelling}

We have used the photoionization code CLOUDY 95.06 (Ferland et al.
1998).

CLOUDY needs as input the information on the shape and intensity of
the radiation from the ionizing source, the chemical composition and
geometry of the nebula, as well as its density and size.  As mentioned
in Section~1, the distance of K~4-47, and thus its size and the
luminosity of the central star, are poorly known.  We adopt the
distance of 5.9~kpc computed by Tajitsu \& Tamura (1998), although
extending the calculations to the full range of distances from 3~kpc
to 7~kpc proposed by Corradi et al.  (2000).

A spherical geometry, and a filling factor of 1.0 have been assumed.
Using the H$\alpha$ and \nii\ images from Corradi et al.~(2000) a core
diameter of 1.9~arcsec was estimated (10\% contour extension corrected
for the finite resolution; Tylenda et al.  2003).  The density was
initially kept to constant and equal to its empirical value
(1900~cm$^{-3}$, see Table~1).  However, as noted in the Introduction,
Aaquist \& Kwok (1990) discovered a bright and very compact radio core
at the centre of K~4-47, implying a high density of
72000~cm$^{-3}$ (assuming an optically thin nebula at 5 GHz,
the distance of 5.9~kpc and following Goudis 1982).  For this reason,
models with much higher density values were also explored.

Dust grains have been included, since they have an important effect,
particularly on the temperature structure of PNe. This effect depends
on the type of grain (graphite and/or silicate), the grain abundances,
and the grain-size distribution, being more relevant in the inner
regions of the nebula (see Dopita \& Sutherland 2000). Both graphite
and silicate grains were considered. As we do not know the grain size
distribution and grain abundances for K~4-47 and more in general for
PNe, we adotped ISM size distribution, with ISM gas-phase depletions
due to grains, following van Hoof et al. (2004).

We adopted a blackbody spectrum for the central star, with \teff\
derived from the \hb, \hei\ and \heii\ emission line ratios, using a
modified Zanstra method as done for instance by Miko{\l}ajewska et al.
(1999).  This gives \teff=130\,000~K (from \heii~4686\AA/\hb) and
\teff=115\,000~K (from \hei~4686\AA/5876\AA).  A lower limit to the
central star luminosity comes from the IRAS spectral energy
distribution (Tajitsu \& Tamura~1998), which gives L$\ge$ 16
D$^2$~$L_{\odot}$ (where D is the distance in kpc) yielding 550
$L_{\odot}$ at 5.9~kpc.

As with the elemental abundances, we first run models with the
empirical abundances given in Table~2, and subsequently with
``average'' values for either normal (Type-II), or Type-I PNe,
following Kingsburgh \& Barlow (1994). 
CLOUDY models with the empirical abundances do not match the observed
line fluxes.  Both helium and \nii~(6548\AA,6583\AA) emission lines
are largely overestimated, while oxygen, sulphur and neon lines are
underestimated, mainly because the very low O/H, Ne/H and S/H input
abundances are not compensated by a high electron temperature in the
model as empirically determined.  Tests with Type-II PNe abundances
were also not successful because the predicted \nii~6548\AA,6583\AA\
intensities are very low, which is expected as the main difference
between Type-I and Type-II abundances is that the latter are depleted
in nitrogen by a factor of $\sim$4 (Kingsburgh \& Barlow 1994).
Assuming Type-I abundances, we found a partial agreement with the
observed spectrum for \teff=120\,000~K, L=550~$L_{\odot}$ and N$_e$=
1900~cm$^{-3}$.  Important optical lines like He{\sc ii}~4686\AA,
\oiii~(4959\AA,5007\AA), \nii~(6548\AA,6584\AA) and
\sii~(6716\AA,6731\AA) are well reproduced by this model, with
discrepancies of 10\% or less. He{\sc i}~4471\AA,~6678\AA, He{\sc
ii}~5412\AA, and [S {\sc iii}]~6312\AA\ lines agree with the observed
spectrum within 12\% to 20\%. Other relatively intense lines such as
\oii~(3726\AA+3729\AA), [Ne{\sc iii}]~3868\AA, and He{\sc i}~5876\AA\
show discrepancies of 35\% to 50\% with respect to the
model. Furthermore, the model underestimates the intensities of the
\oiii~4363\AA\ and the \nii~5755\AA\ lines by a factor of about 3.
These lines are the keys for the determination of the electron
temperature. Finally, the model underestimates [N{\sc i}]~5200\AA\ and
[O{\sc i}]~6300\AA\ by factors larger than 30.  Let us note, in
addition, that this model requires a size for the core of 4.0~arcsec,
twice the optical size observed.

It is then instructive to explore alternative scenarios. Let us
first note that in some PNe
and related objects (see e.g. Corradi 1995), high \oiii~4363\AA\ and
\nii~5755\AA\ line intensities have been interpreted as the
signature of very high core densities, as at $N_e$ larger than about
10$^5$~cm$^{-3}$ the auroral (\oiii~4363\AA\ and \nii~5755\AA) to
nebular (\oiii~4959\AA\ and \nii~6583\AA) line ratios are indicators
of density rather than of temperature (Gurzadyan 1970).  For this
reason, we have calculated other models assuming much higher (albeit
still constant) densities.  As noted above, high core densities are
implied by the radio size and flux (Aaquist \& Kwok 1990).

CLOUDY models for the observed radio core size, 0.25~arcsec, and such
high densities (from 72000~K up to $3.0\times10^5$cm$^{-3}$) show that
both \oiii~4363\AA\ and \nii~5755\AA\ intensities can be now
reproduced, but other nebular lines, in particular
\oii~(3726\AA+3729\AA), \nii~(6548\AA,6583\AA), and
\sii~(6717\AA,6731\AA), become now largely underestimated because of
collisional quenching.  A natural way to solve the problem might be to
assume a strong density stratification in the core of K~4-47, with a
very dense inner zone where \oiii~4363\AA\ and \nii~5755\AA\ are
mostly formed, and a lower density outer region where other important
nebular lines are produced.  This idea should be tested by means of an
appropriate photoionization model; 3-D codes (like MOCASSIN; Ercolano
et al.~ 2003) are much better suited than CLOUDY to deal with such
extreme density variations.

In summary, we find that none of the constant density models is able
to account, simultaneouly, for all optical emission lines in the
\co.  In particular, the \oiii~4363\AA\ and \nii~5755\AA\ intensities
are strongly understimated if the nebular density is the one derived
empirically from the \sii\ lines.  A model with a strong density
stratification could possibly offer a solution to the problem.

\section[]{Shock modelling for the knots}

The fact that we are not able to provide an accurate representation of
the \co\ spectrum, also prevents us to determine the radiation
escaping from the core and reaching the knots, and thus to attempt a
reliable photoionization mo\-del\-ling of the latter. A consistent
photoionization modelling of the whole nebula (core+knots) would also
require a fully 3-D modelling, that goes beyond the scopes of the
present work. Given the evidence that the knots might be excited by
shocks (Section.~3.2), we instead attempt to describe their spectrum
using existing shock models.

\subsection{Bow-Shock Models}

We have used the bow-shock models described in Raga \& Bohm (1986) and
Hartigan et al. (1987).  We consider a bow shock with a functional
form, $z/a= (r/a)^p$, where $z$ is measured along the symmetry axis,
$r$ is the cylindrical radius, the parameter $p$ determines the shape
of the bow shock and the constant $a$ determines its size (Beck et al.
2004).  The emission from the bow-shock is modeled as described by
Hartigan et al.  (1987), being the geometry of the bow shock the
parameter that determines the shock velocity.

\begin{figure}
\vbox{
\psfig{file=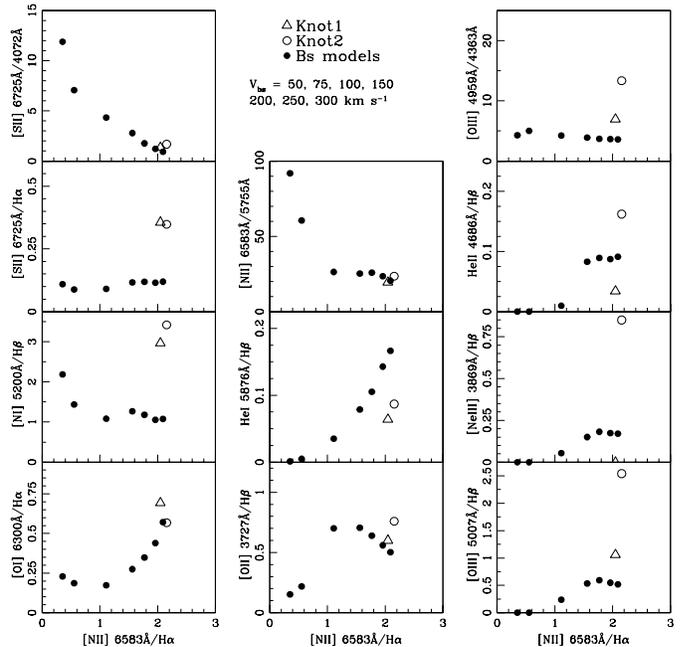,width=9.0truecm,bbllx=36pt,bblly=165pt,bburx=585pt,bbury=690pt}}
\caption{Results of the bow-shock models, in terms of line ratios, as
compared to the observed emission. Results are shown for bow-shock
velocities of V$_{bs}$ of 50, 75, 100, 150, 200, 250, and 300~\kms,
corresponding to points moving from left to the right side inside each
panel.}
\label{bshock}
\end{figure}

The plane-paralel shock models required to predict the bow-shock
emission line ratios are obtained with the photoionization-shock code
MAPPINGS Ic.  We have adopted a pre-shock density of 400 cm$^{-3}$,
that is within the range generally assumed for stellar jets.  The
pre-shock magnetic field was taken to be negligible.  We assumed
ele\-men\-tal abundances for He, C, N, O, Ne, S, and Ar to be within the
range of Type-I PN values. As for the preionization of the gas, local
equilibrium is assumed. We consider the bow-shock velocity as a free
parameter, adopting a set of values from 50 to 300 km s$^{-1}$.

\subsection{Comparison with the Observed Spectra}

We have obtained the best fit to the observed spectra for $p$=3.  A
summary of the results is presented in Figure~\ref{bshock} for a
number of emission line ratios plotted as a function of
\nii~6583\AA/\ha.  Note that the later line ratio increases in
strength as the shock velocity increases, and in the case of K~4-47 is
matched by only the largest shock velocities considered, somewhat
larger than the estimate of 150~\kms\ by Corradi et al.~(2000).

For these velocities, the bow-shock models fit pretty well the line
ratios \sii~(4069\AA+4076\AA)/(6717\AA+6731\AA)$\equiv$\sii~4072\AA/6725\AA,
and \nii~6583\AA/5755\AA, sensitive to the temperature in the
recombination zone where \sii\ and \nii\ are formed, but not equally
well the \oiii~4959\AA/4363\AA\ ratio, indicating that the predicted
electron temperature \te\oiii\ is too large.  The \nii\ and \oiii\
electron temperatures predicted by these models, for both knots, are
17800~K and 27600~K, respectively.

The ratios \oi~6300\AA/\ha\ and \oii~(3726\AA+3729\AA)/\hb\ are also
well reproduced.  Larger discrepancies appear for the helium lines
He{\sc i}~5876\AA\ and He{\sc ii}~4686\AA, and specially for [Ne {\sc
iii}]~3868\AA\ and \oiii~5007\AA\ in \kii, and \sii~(6716\AA+6731\AA)
and [N {\sc i}]~5200\AA\ in both knots.  Note that a similar problem
for the \sii~(6717\AA+6731\AA)/\ha\ ratio is also found in HH objects
(Raga et al.  1996).  The large discrepancy for the [N {\sc
i}]~5200\AA/\hb\ emission line ratio is intriguing, as strong line
intensities for this ion can only be produced with low shock
velocities, which could however not account for the observed flux of
higher ionization stages like \nii~6583\AA. Different shock
geometries, producing a different distribution of shock velocities,
might help to solve this problem.

Also note that the bow-shock models are based on appraisal of what
would be reasonable parameters for the knots of K~4-47, and we did not
intend to reproduce the line intensities separately for each of the
knots.  But the knots are not identical: in particular, high excitation
lines like He{\sc ii}~4686\AA, [Ne {\sc iii}]~3868\AA\ and
\oiii~5007\AA\ are much brighter in \kii\ than in \ki, which
indicates that the radiation field of the central star
would affect the \kii\ emission line spectrum much more than that of \ki.
The observed spectra of \kii\ would also fit a model with a
somewhat larger N/H abundance and a lower bow shock velocity.

Summarising, the analysis of the shock-excited knots suggests that
they are high-velocity (250 to 300\kms) condensations with high
electron temperatures (\te\nii\ =17800~K) and chemical abundances
similar to Type-I PNe.

\section{Discussion}

The analysis presented in this paper is the first attempt to determine
the physical parameters, excitation mechanism and chemical abundances
of K~4-47.  Among other PNe with LISs, M~2-48 (L\'opez-Mart\'\i n et
al.  2002) and Kj~Pn~8 (L\'opez, V\'azquez \& Rodr\'\i guez  1995) have, 
similarly to
K~4-47, a high electron temperature and the evidence for shock-excited
LISs.  But note that in these PNe, except for one of the M~2-48's
knots (for which \te\ is 20100~K, twice the core temperature), \te\
and \ne\ are not estimated in a detailed zone-by-zone basis, because
\oiii~4363\AA\ and \nii~5755\AA\ were not measured.

The two distinct models that we used for the analysis of K~4-47 (pure
photoionization for the core and pure shock-excitation for the pair of
knots), provide some clues for the interpretation of this nebula.

First, the analysis of the core would suggests He and N overabundances
typical of Type-I (Galactic disk) PNe, but also an extreme oxygen
deficiency which is instead more ty\-pi\-cal of PNe in the Galactic halo.
These abundances calculation would however be invalidated if more
detailed modelling proves the existence of a strong density
stratification within the core, with an inner region with extremely
high density ($\sim 10^5$~cm$^{-3}$), and an outer zone with a much
lower density matching the empirical determination from the \sii\
lines.  In fact, in this case the \te\ determined empirically using
the \oiii~4959\AA/4363\AA\ and \nii~6583\AA/5755\AA\ line ratios would
be wrong, as the auroral and nebular lines involved would be formed in
different regions.  Note that, albeit unusual, such high density cores
were actually found in several other PNe (Corradi 1995), which like
K~4-47 show an extreme degree of collimation.

Second, the knots seem to be mainly shock-excited.  This conclusion is
strenghtened by diagnostic line ratio diagrams, and,
interestingly, by the fact that the H$_2$ emission of K~4-47 is also
shock-excited, contrary to most PNe where instead it is excited by
fluorescence (Lumsden et al.~2001).  Our shock modelling also
indicates that knots have Type-I abundances.  According to these
models, knots would move with space velocities of 250-300\kms, about
twice the value estimated by Corradi et al.  (2000) from the analysis
of the lines profiles.  Moreover, while the shock model is successful
in matching the \nii~(6583\AA/5755\AA) line ratio at the knots it
un\-de\-res\-ti\-ma\-te their \oiii~(4959\AA/4363\AA) ratio by some 30\% and 70\%
for \ki\ and \kii, respectively.

As a final remark, the two physical processes explored
separately in this paper, i.e.  photoionization and shock heating, are
likely to be simultaneously present in a real nebula like K~4-47, as
gas can be ionized by both energetic photons from the central star and
from the shocks associated with the observed supersonic
outflows.  Taking this into account would require a
sophisticated modelling which is not presently available, and which
would also need to be supported by a better knowledge of the nebular
and stellar parameters.  In particular, very little information is
presently known about the distance and the properties of the central
star of K~4-47, a basic data that could be addressed by future
observations.

\section*{Acknowledgments}

We thank the anonymous referee for his/her comments, which helped us to 
improve the paper significantly. We also thank Joanna Miko{\l}ajewska 
for fruitful discussions and Corrado Giammanco for helping with CLOUDY.  
The work of DRG, AM, RLMC and LLM is partially supported by a grant from 
the Spanish Ministry of Science and Technology (AYA 2001-1646). The work 
of AR is supported by the grant AYA 2002-00205.

\begin{table*}
\scriptsize{
\centering
 \begin{minipage}{80mm}
 \caption{Observed emission line fluxes (normalized to \hb\ =100),
 electron density and temperatures. The estimated percentage errors in
 the line fluxes (for a range of fluxes relative to \hb) are given at
 the bottom. }
 \begin{tabular}{@{}llll@{}}
 \hline
& \multicolumn{3}{c}{Line Fluxes} \\
Line Identification (\AA) & \ki\ & \co\ & \kii\ \\
 \hline
{}[O{\sc ii}] 3726.0 + 3728.8                    &23.4  &26.6  &33.3   \\
H9  3835.4                                      &9.45  &-          &-  
       \\
{}[Ne{\sc iii}] 3868.7                            &-     &48.9  &40.3   \\
He{\sc i} 3888.7 + H8 3889.1                    &-     &7.05  &12.8   \\
{}[Ne{\sc iii}] 3967.5 + H$\epsilon$ 3970.1 &11.5  &38.9  &27.8   \\
{}[S{\sc ii}] 4068.6                             &23.1  &3.18  &20.0   \\
{}[S{\sc ii}] 4076.4                             &15.0  &2.49  &10.7   \\
H$\delta$ 4101.8                             &15.6  &16.1  &16.6   \\
H$\gamma$ 4340.5                             &30.1  &32.0  &27.8   \\
{}[O{\sc iii}] 4363.2                             &3.45  &20.8  &4.56   \\
He{\sc i} 4471.5                             &-     &4.28  &9.22   \\
N{\sc iii} + O{\sc ii} 4641.0                     &3.69  &4.40  &-      
   \\
He{\sc ii} 4685.7                             &2.95  &24.4  &14.3   \\
H$\beta$ 4861.3                             &100   &100   &100    \\
{}[O{\sc iii}] 4958.9                             &35.9  &361   &100    \\
{}[O{\sc iii}] 5006.86                             &118   &1088  &279    \\
{}[N{\sc i}] 5200.2                             &367   &42.0  &421    \\
He{\sc ii} 5411.6                             &-     &4.11  &-          \\
{}[O{\sc i}] 5577.4                            &12.6  &-          &10.4   \\
{}[N{\sc ii}] 5754.6                             &51.4  &43.6  &40.3   \\
He{\sc i} 5875.7                             &11.5  &42.5  &14.6   \\
He{\sc ii} 5882.0                            &-     &-          &5.54   \\
He{\sc ii} 5896.5                            &5.82  &-          &10.1   \\
{}[O{\sc i}] 6300.3                             &427   &54.7  &305    \\
{}[S{\sc iii}] 6312.1                             &-     &7.65  &-      
   \\
{}[O{\sc i}] 6363.8                              &146   &17.1  &99.1   \\
{}[N{\sc ii}] 6548.0                              &446   &333   &415.2  \\
H$\alpha$ 6562.8                             &681   &762   &587.8  \\
{}[N{\sc ii}] 6583.4                              &1406  &976   &1276   \\
He{\sc i} 6678.1                              &3.22  &13.4  &-          \\
{}[S{\sc ii}] 6716.5                              &93.5  &32.1  &86.0   \\
{}[S{\sc ii}] 6730.8                              &165   &45.2  &130    \\
\\
N$_e$[S{\sc ii}](cm$^{-3}$)     &  4600$\pm$850  &  1900$\pm$420  &
 2400$\pm$400  \\
T$_e$[O{\sc iii}](K)            & $\ge$21000     & 19300$\pm$2300 &
16100$\pm$4400 \\
T$_e$[N{\sc ii}](K)             & 18900$\pm$2950 & $\ge$21000     &
16950$\pm$2800 \\
\\
\hline
& \multicolumn{3}{c}{Percentage errors in line fluxes} \\
Line Fluxes  & \ki\ & \co\ & \kii\ \\
\hline
(0.01--0.05)I$_{{\rm H}\beta}$  &51   &41  &58 \\
(0.05--0.15)I$_{{\rm H}\beta}$  &23   &15  &26 \\
(0.15--0.30)I$_{{\rm H}\beta}$  &14   &11  &15 \\
(0.30--2.0)I$_{{\rm H}\beta}$   &12   &8   &11 \\
(2.0--5.0)I$_{{\rm H}\beta}$    &12   &7   &9  \\
(5.0--10.0)I$_{{\rm H}\beta}$   &11   &7   &9  \\
$>$ 10 I$_{{\rm H}\beta}$       &11   &7   &9  \\
\hline
\end{tabular}
\end{minipage}
}
\end{table*}

\begin{table*}
\scriptsize{
\centering
 \begin{minipage}{43mm}
 \caption{Ionic and total abundances of the Core.
Percentage errors are given within brackets.}
 \begin{tabular}{@{}ll@{}}
 \hline
\multicolumn{1}{c}{Elemen/Iont} \ \ \ \ &  \multicolumn{1}{c}{Abundances} \\
 \hline
 He$^+$/H      &\ \ \ \   1.14E-01(08)     \\
 He$^{2+}$/H   &\ \ \ \   2.56E-02(14)     \\
 {\bf He/H }        &\ \ \ \  {\bf 1.39E-01(14)} \\
 O$^0$/H       &\ \ \ \   5.10E-06(10)     \\
 O$^+$/H       &\ \ \ \   3.09E-06(14)     \\
 O$^{2+}$/H    &\ \ \ \   6.13E-05(21)     \\
 {\it icf}(O)        &\ \ \ \   1.14            \\        
 {\bf O/H}        &\ \ \ \   {\bf 7.37E-05(32)} \\
 N$^0$/H       & \ \ \ \  8.47E-06(14)     \\
 N$^+$/H       & \ \ \ \  1.57E-05(8)           \\
 {\it icf}(N)        &\ \ \ \   23.8            \\
 {\bf N/H}        &\ \ \ \   {\bf 3.74E-04(40)} \\
 Ne$^{2+}$/H   &\ \ \ \   1.44E-05(32)     \\
 {\it icf}(Ne)  &\ \ \ \   1.20                    \\
 {\bf Ne/H}        &\ \ \ \   {\bf 1.74E-05(66)} \\
 S$^+$/H       & \ \ \ \  2.24E-07(14)     \\
 S$^{2+}$/H    &\ \ \ \   7.44E-07(36)     \\
 {\it icf}(S)        &\ \ \ \   2.02            \\
 {\bf S/H}        &\ \ \ \   {\bf 1.96E-06(48)} \\
\hline
\end{tabular}
\end{minipage}
}
\end{table*}

\bsp
\label{lastpage}
\end{document}